\documentclass[aps,pra,twoside,centertags,nofootinbib,eqsecnum]{revtex4}
 \usepackage{amsmath,amssymb,bm,euscript}
 \usepackage{graphicx,boxedminipage}
 \usepackage[bf,small]{caption2}
 \usepackage{floatflt}
 \usepackage[hypertex]{hyperref}
 \usepackage{xcolor,fancybox}

\definecolor{Dark-blue}{RGB}{0,0,255}
\newcommand{\bra}[1]{\ensuremath{\bm{\langle}#1\bm{|}}}
\newcommand{\ket}[1]{\ensuremath{\bm{|}#1\bm{\rangle}}}
\newcommand{\Tr}{\ensuremath\mathrm{Tr\,}}

\reversemarginpar \marginparwidth=1,2cm

\begin{document}

\title{The spectrum of non-centrosymmetrically layered spherical cavity resonator.\\I. The mode decomposition method}

\author{Z.\,E. Eremenko, Yu.\,V. Tarasov, and I.\,N. Volovichev}

\affiliation{O.\,Ya. Usikov Institute for Radiophysics and Electronics, NAS of Ukraine,\\
12 Proskura Str., Kharkiv 61085, Ukraine}

\begin{abstract}
  We develop a theoretical method for solving Maxwell's equations to obtain the frequency spectra of inhomogeneous and asymmetric cavity resonators using a couple of effective Debye-type potentials. The structure we study specifically is the layered spherical cavity resonator with symmetrically or asymmetrically inserted inner dielectric sphere. The comparison of the exact numerical results obtained for the frequency spectrum of layered cavity resonator with centrosymmetrically inserted sphere and the spectrum found from the suggested theory reveals good agreement at the initial part of the frequency axis. The coincidence accuracy depends on the number of trial resonant modes that we use while approving our method numerically.
\end{abstract}

\maketitle

\allowdisplaybreaks
\section*{INTRODUCTION}
\label{Intro}

The problem of chaos in the spectra of wave (or, equivalently, quantum) systems has attracted a great deal of researchers' attention for many years in view of both its mathematical non-triviality and the applied significance. The solution to specific problems arising in the framework of this versatile problem relates closely to the concepts and the developments in the theory of chaos of classical dynamical systems \cite{bib:Heiss92}. For wave systems of macroscopic and mesoscopic dimensions, in particular, for quasi-optical electromagnetic resonators, the main tool for investigating their chaotic properties, just the same as for dynamical systems, is the statistical analysis of their spectra \cite{bib:Stockman99}.

Most of the conclusions of the statistical theory regarding the dynamical system spectra settle on their symmetry properties \cite{bib:BohigGian75,bib:BohigGianSchmit84}, since direct calculation of the spectral parameters is normally difficult to perform. Similar methods apply to wave systems as well, and conclusions about their chaotic properties are basically made basing on the ray ("trajectory") optics \cite{bib:BarYanNaydKur06}. This approach, in view of significant uncertainty of the position of the indicative points on the ray trajectories, does not provide sufficient reliability of the results, especially when it comes to closed systems such as cavity resonators. Meanwhile, to obtain the reliable spectrum for resonators with inhomogeneities in the bulk appears to be, as a rule, the problem extremely expensive computationally, if one takes account of the number of (Maxwell's) equations subject to the analysis.

The latter circumstance was previously overcome only for homogeneous systems through the introduction in them of so-called Debye potentials \cite{bib:Stratton07,bib:Vainstein88}. However, although this approach proved to be quite productive, it was not possible for a long time to extend it to heterogeneous systems as well. One of the few successful attempts to do this can be noted in Ref.~\cite{bib:Malykh_etal_17}, where a method for solving the Helmholtz equation in inhomogeneous closed waveguides was developed by reducing it to the Hamiltonian form with the use of \emph{four} scalar potentials.

In this paper, we propose a method for calculating the spectrum of inhomogeneous and asymmetric cavity resonators by introducing \emph{a couple} of Debye-type effective potentials, with the subsequent solution of the emerging wave equations in the whole resonator rather than in its partial regions using the method of separation of the oscillation modes. The separation, even for heterogeneous systems, is a strict procedure, in contrast to the approximate separation normally performed within the framework of the conventional coupled-mode theory \cite{bib:HausHuang91}.

The suggested mode decomposition method involves the solution of matrix equations either to find their eigenvalues or to calculate the Green function. In both cases, it is impossible to find the closed analytical solution, and it is necessary to apply also some numerical methods. For this, in its turn, it is necessary to deliberate some issues. In the theory set forth below, the expansion of the Debye scalar potentials is carried out over an infinite orthonormal set of basis functions. Obviously, when the computer is used, one has to limit himself to a finite number of such functions. It requires, for one thing, to determine the sensitivity of the spectrum to the dimension of the basis set. Secondly, it is necessary to analyze the accuracy and stability of the computational scheme when finding the matrix elements of the potentials. Finally, the calculation of the spectrum in a sufficiently large frequency interval implies quite a~lot of processor time. So, it is important to choose the efficient computational algorithm, in particular, for numerical integration in 3D. We present here our results on this subjects as well.

\section{STATEMENT OF THE PROBLEM AND BASIC EQUATIONS}
\label{StateProb}

\begin{figure}[h]
  \setcaptionmargin{.5in}%
  \centering
  \scalebox{.6}[.6]{\includegraphics{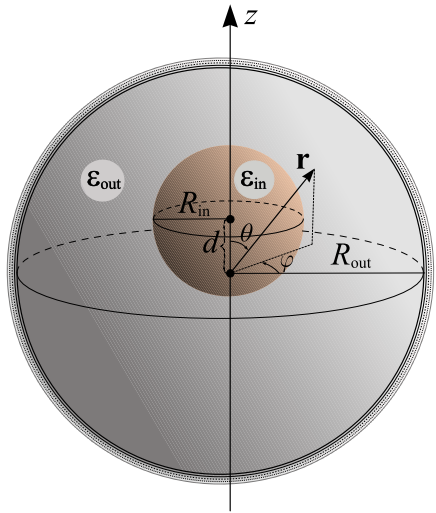}}
 \caption{The layered spherical cavity resonator.}
 \label{DoubleSpherRes}
\end{figure}

The goal of this paper is to determine the degree of chaos in the spectrum of oscillations of electromagnetic (EM) cavity resonator of radius $R_{\mathrm{out}}$ filled with a piecewise homogeneous dielectric, as shown in Fig.~\ref{DoubleSpherRes}. The permittivity of the outer dielectric layer is $\varepsilon_{\mathrm{out}}$
whereas of the inner spherical inclusion of radius $R_{\mathrm{in}}$, whose center is shifted to arbitrary distance $d$ from the center of outer sphere, is equal to $\varepsilon_{\mathrm{in}}$. The magnetic permeabilities of both parts of the resonator infill are assumed to be the same, and we put them equal to unity in the final formulas.

Normally, to find the EM field strength in a system with heterogeneously distributed electric and magnetic parameters it is required to solve the set of Maxwell equations, which in the stationary case is represented by the following eight coupled scalar equalities,
\begin{subequations}\label{Maxwell_EH}
\begin{align}
\label{rotE}
 & \mathrm{rot}\,\mathbf{E}=ik\,\mu(\mathbf{r})\mathbf{H}\ ,\\
\label{rotH}
 & \mathrm{rot}\,\mathbf{H}=-ik\,\varepsilon(\mathbf{r})\mathbf{E}\ ,\\
\label{divB}
 & \mathrm{div}\big[\mu(\mathbf{r})\mathbf{H}\big]=0\ ,\\
\label{divD}
 & \mathrm{div}\big[\varepsilon(\mathbf{r})\mathbf{E}\big]=0\ .
\end{align}
\end{subequations}
Here, $k=\omega/c$, $\mu(\mathbf{r})$ and $\varepsilon(\mathbf{r})$ are the permeability and the permittivity of the resonator infill, which are assumed to be the functions of the coordinate vector $\mathbf{r}$.

The number of equations subject to solution may be reduced if one goes over from the vector-valued electric and magnetic fields to the scalar-valued Debye potentials. Until now, these potentials have been consistently introduced only for homogeneous systems \cite{bib:Stratton07,bib:Vainstein88}. In this paper, we show that similar potentials can be established in heterogeneous systems as well, and without any special restrictions on the degree and the character of inhomogeneity.

Following monograph \cite{bib:Vainstein88}, we represent the fields $\mathbf{E}$ and $\mathbf{H}$ in the form of the sums of the components of ``electric'' and ``magnetic'' type,
\begin{subequations}\label{EH=electr+magn}
\begin{align}
\label{E=electr+magn}
  \mathbf{E} & =\mathbf{E}^e+\mathbf{E}^m\ ,\\
\label{H=electr+magn}
  \mathbf{H} & =\mathbf{H}^e+\mathbf{H}^m\ .
\end{align}
\end{subequations}
The electric-type oscillations (\emph{e}) are determined by the vanishing of the radial component of their magnetic field, $H_r^e\equiv 0$, whereas the oscillations of magnetic type (\emph{m}) are determined by the equality $E_r^e\equiv 0$. In accordance with the above definitions, the first type of oscillation is also called TM, and the second type is called TE oscillations in a~spherical resonator. In the system we study, instead of the vector fields $\mathbf{E}$ and $\mathbf{H}$ (both of electric and magnetic types) we will use two scalar functions $U(\mathbf{r})$ and $V(\mathbf{r})$ (commonly called Hertz \emph{functions}) through which all the EM field components can be unambiguously expressed. The method of introducing these functions in the case of inhomogeneous systems is quite sophisticated, and we briefly describe it in the following way.

A pair of equations \eqref{divB} and \eqref{divD} can be fulfilled identically through the proper choice of the calibration of the conventional scalar and vector potentials. In order to satisfy also equations \eqref{rotE} and \eqref{rotH}, in the case of \emph{e}-fields one can make substitution
\begin{equation}\label{Ephi_Etheta_Er->U}
 \begin{aligned}
 & E_{\varphi}^e=\frac{1}{\varepsilon(\mathbf{r})\,r\sin\vartheta}\cdot\frac{\partial^2 U} {\partial\varphi\partial r}\ , \qquad\quad
  E_{\vartheta}^e=\frac{1}{\varepsilon(\mathbf{r})\,r}\cdot
  \frac{\partial^2U}{\partial\vartheta\partial r}\ ,\\
 &  E_r^e=-\frac{1}{\varepsilon(\mathbf{r})\,r^2\sin\vartheta}
  \left[\frac{\partial}{\partial\vartheta}\left(\sin\vartheta\frac{\partial U} {\partial\vartheta}\right)+ \frac{1}{\sin\vartheta}\frac{\partial^2U}{\partial\varphi^2}\right]\ ,\\[6pt]
 & H_{\varphi}^e=\frac{ik}{r}\frac{\partial U}{\partial\vartheta}\ ,\qquad\qquad
  H_{\vartheta}^e=\frac{-ik}{r\sin\vartheta}\frac{\partial U}{\partial\varphi}\ .
 \end{aligned}
\end{equation}
A couple of equalities \eqref{rotE} and \eqref{rotH} in this case is met if the following two equations are simultaneously satisfied, viz.,
\begin{subequations}\label{Two_eqs_for_U-full}
\begin{align}\label{1st_eq_for_U-full}
 & \frac{\partial}{\partial r}\left[\frac{1}{\varepsilon(\mathbf{r})}
  \frac{\partial^2U}{\partial\varphi\partial r}\right]+
  \frac{1}{r^2}\frac{\partial}{\partial\varphi}
  \left\{\frac{1}{\varepsilon(\mathbf{r})}
  \left[\frac{1}{\sin\vartheta}\frac{\partial}{\partial\vartheta}
  \left(\sin\vartheta\frac{\partial U}{\partial\vartheta}\right)+
  \frac{1}{\sin^2\vartheta}\frac{\partial^2U}{\partial\varphi^2}\right]\right\}+
  k^2\mu(\mathbf{r})\frac{\partial U}{\partial\varphi}=0
%
\intertext{and}
%
\label{2nd_eq_for_U-full}
 & \frac{\partial}{\partial r}\left[\frac{1}{\varepsilon(\mathbf{r})}
  \frac{\partial^2U}{\partial\vartheta\partial r}\right]+
  \frac{1}{r^2}\frac{\partial}{\partial\vartheta}
  \left\{\frac{1}{\varepsilon(\mathbf{r})}
  \left[\frac{1}{\sin\vartheta}\frac{\partial}{\partial\vartheta}
  \left(\sin\vartheta\frac{\partial}{\partial\vartheta}\right)+
  \frac{1}{\sin^2\vartheta}\frac{\partial^2}{\partial\varphi^2}\right]U\right\}+
  k^2\mu(\mathbf{r})\frac{\partial U}{\partial\vartheta}=0\ .
\end{align}
\end{subequations}
For the magnetic-type fields, substitutions analogous to \eqref{Ephi_Etheta_Er->U}, but with the interchange $E\rightleftarrows H$, $U\rightarrow V$ and $\varepsilon(\mathbf{r})\rightleftarrows\mu(\mathbf{r})$ result in equations that differ from Eqs.~\eqref{Two_eqs_for_U-full} by the same substitution.

Hertz functions $U(\mathbf{r})$ and $V(\mathbf{r})$ are not so convenient for representing the fields in the resonator with them. More convenient appear to be the functions that are related to the above functions by the equalities $u(\mathbf{r})=U(\mathbf{r})/r$ and $v(\mathbf{r})= V(\mathbf{r})/r$. These functions are commonly termed the Debye potentials (DPs) \cite{bib:Vainstein88}. The equations for them are convenient in that they allow one to seek the solution in the form of an expansion in Fourier series with respect to the well-practical spherical functions, and therefore we shall continue the exposition in terms of precisely these potentials.

\section{SEPARATION OF TM AND TE OSCILLATIONS}
\label{Separation}

Before proceeding directly with the solution of the equations for the DPs we dwell on the boundary conditions for these potentials. The violation of the central symmetry of the resonator in the case of a nonzero displacement of its internal spherical insert ($d\neq 0$) should, at first glance, result in the impossibility to separate TM and TE oscillations. Yet, within the framework of the formalism we propose in our study these oscillations are still exactly separable. Indeed, the boundary condition for the electric field tangential components to vanish on the boundary of the outer sphere, being written in terms of the DPs, yields the following two equalities,
\begin{subequations}\label{BC(UV)}
   \begin{align}\label{BC(E_theta)}
  &  \left[\frac{1}{\varepsilon_{\text{out}}\,r}\frac{\partial}{\partial\vartheta}
    \left(u+r\frac{\partial u}{\partial r}\right)+
    \frac{ik}{\sin\vartheta}\frac{\partial v}{\partial\varphi}\right]
    \Bigg|_{r=R_{\text{out}}}=0\ ,\\
\label{BC(E_phi)}
  &  \left[\frac{1}{\varepsilon_{\text{out}}\,r\sin\vartheta}\frac{\partial}{\partial\varphi}
    \left(u+r\frac{\partial u}{\partial r}\right)-
    ik\frac{\partial v}{\partial\vartheta}\right]
    \Bigg|_{r=R_{\text{out}}}=0\ .
   \end{align}
\end{subequations}
They are equivalent to a pair of more simple relations, which do not connect potentials $u(\mathbf{r})$ and $v(\mathbf{r})$ with each other, namely,
\begin{subequations}\label{BC(u|v)}
 \begin{align}\label{BC(Psi_U)}
&  \left(u+r\frac{\partial u}{\partial r}\right)\bigg|_{r=R_{\text{out}}}=0 \ ,\\[6pt]
\label{BC(Psi_V)}
&\  v(r=R_{\text{out}})=0 \ .
 \end{align}
\end{subequations}
The choice of the boundary conditions in the form of equalities \eqref{BC(u|v)} allows one to seek the electric- and magnetic-type fields in the resonator with broken symmetry independently of each other, thus representing the oscillations of a general type in the form of a superposition of unrelated TM and TE oscillations.

\section{THE DECOMPOSITION OF THE RESONATOR MODES}
\label{ModeDecomp}

It is convenient to proceed with direct solution of the problem not in the coordinate but rather in oscillation mode representation, by Fourier transforming the equations for Debye potentials with respect to the appropriate complete set of eigen-functions. As such a set, we take the normalized eigen-functions of the Laplace operator for a homogeneous sphere of radius $R_{\mathrm{out}}$, which in spherical coordinates have the form
\begin{align}\label{basis_funcs}
 & \ket{\mathbf{r};\bm{\mu}}=\frac{D_n^{(l)}}{R_{\mathrm{out}}}\sqrt{\frac{2}{r}}\,
  J_{l+\frac{1}{2}}\left(\lambda_n^{(l)}r/R_{\mathrm{out}}\right)Y_l^m(\vartheta,\varphi)\\
  \notag
 & \big(n=1,  2,\ldots,\infty\ ;\quad l=0,1,2,\ldots,\infty\ ;\quad  m=-l,-l+1,\ldots l-1,\ l\big)\ .
\end{align}
Here $\bm{\mu}=\{n,l,m\}$ is the vector mode index whose components correspond to the radial, polar, and azimuthal coordinate variables, respectively; $D_n^{(l)}$ is the normalization coefficient, $J_{p}\big(u)$ is the Bessel function of the first kind, $Y_l^m(\vartheta,\varphi)$ is the spherical function; $\lambda_n^{(l)}$ are the positive zeros either of the sum $J_{l+\frac{1}{2}}(u)+2uJ'_{l+\frac{1}{2}}(u)$, if boundary condition \eqref{BC(Psi_U)} is used, or of the function $J_{l+\frac{1}{2}}(u)$ in the case of boundary condition \eqref{BC(Psi_V)}. The eigenvalues of the Laplace operator which correspond to functions \eqref{basis_funcs} are equal to
\begin{equation}\label{Eigen-energies}
  E_{\bm{\mu}}=-k_{\bm{\mu}}^2 =-\left(\frac{\lambda_n^{(l)}}{R_{\mathrm{out}}}\right)^2\ .
\end{equation}
They are degenerate in azimuth index $m$ with the degeneracy order of $2l+1$.

In the mode variables the equations for Debye potentials, given the conditions \eqref{BC(u|v)}, turn into two independent pairs of interrelated equations for the Fourier components of these potentials. For nonmagnetic materials ($\mu(\mathbf{r})\equiv 1$), equations \eqref{Two_eqs_for_U-full}, being Fourier transformed at first over angle $\varphi$, acquire the following form:
\begin{subequations}\label{u_eqs-azimuth_m}
\begin{align}
\label{u_eq-azimuth_m}
  m &\left[ \hat{\Delta}_m+
  k^2\overline{\varepsilon}-\hat{V}^{(\varepsilon)}-\hat{V}^{(r)}\right]u_m(r,\vartheta)=0
  \ ,\\[6pt]
\label{u_m-eq2}
 & \left[ \hat{\Delta}_m+
  k^2\overline{\varepsilon}-\hat{V}^{(\varepsilon)}-\hat{V}^{(r)}\right]
  \frac{\partial u_m(r,\vartheta)}{\partial\vartheta}-\hat{W}_m^{(\vartheta)}u_m(r,\vartheta)=0\ .
\end{align}
\end{subequations}
In unison, the equations for Fourier components of the potential $v(\mathbf{r})$ are as follows,
\begin{subequations}\label{v_eqs-azimuth_m}
\begin{align}
\label{u_eq-azimuth_m}
  m &\left[ \hat{\Delta}_m+
  k^2\overline{\varepsilon}-\hat{V}^{(\varepsilon)}\right]v_m(r,\vartheta)=0
  \ ,\\[6pt]
\label{v_m-eq2}
 & \left[ \hat{\Delta}_m+
  k^2\overline{\varepsilon}-\hat{V}^{(\varepsilon)}\right]
  \frac{\partial v_m(r,\vartheta)}{\partial\vartheta}=0\ .
\end{align}
\end{subequations}
In Eqs.~\eqref{u_eqs-azimuth_m} and \eqref{v_eqs-azimuth_m}, $\overline{\varepsilon}$ denotes the bulk-averaged dielectric constant of the substance that fills up the resonator,
\begin{align}
\label{AvEps_Sigma}
  \overline{\varepsilon} =
  \varepsilon_{\mathrm{out}}+
  (\varepsilon_{\mathrm{in}}-\varepsilon_{\mathrm{out}})
  \left(R_{\mathrm{in}}/R_{\mathrm{out}}\right)^3\ ,
\end{align}
$\hat{\Delta}_m$ is the Laplace operator for $m$-th azimuth mode, $\hat{V}^{(\varepsilon)}$ and $\hat{V}^{(r)}$ are effective ``potentials'' possessing the operator structure, viz.,
\begin{equation}\label{V^(e)&V^(r)}
  \hat{V}^{(\varepsilon)}=-k^2\Delta\varepsilon(r,\vartheta)\ ,\qquad
  \hat{V}^{(r)}=\frac{\partial\ln\varepsilon(r,\vartheta)}{\partial r}
  \left(\frac{1}{r}+ \frac{\partial}{\partial r} \right)\ ,
\end{equation}
$\Delta\varepsilon(r,\vartheta)=\varepsilon(r,\vartheta)- \overline{\varepsilon}$ , and $\hat{W}_m^{(\vartheta)}$ is the operator of the following form:
\begin{equation}\label{V^(theta)-def}
  \hat{W}_m^{(\vartheta)}=\frac{\partial\ln\varepsilon(r,\vartheta)}{\partial\vartheta}\cdot
  \frac{1}{r^2}\left[\frac{1}{\sin\vartheta}\frac{\partial}{\partial\vartheta}
  \left(\sin\vartheta\frac{\partial}{\partial\vartheta}\right)-
  \frac{m^2}{\sin^2\vartheta}\right]\ .
\end{equation}

Both equation pairs, \eqref{u_eqs-azimuth_m} and \eqref{v_eqs-azimuth_m}, in view of the separability of TM (electric) and TE (magnetic) oscillations, must be consistent independently of each other. If the set \eqref{v_eqs-azimuth_m} is consistent on the spectrum of the operator entering square brackets for any azimuth mode index, then the set \eqref{u_eqs-azimuth_m} is such for the mode with $m=0$ only. This means that in the magnetically homogeneous resonator depicted in Fig.~\ref{DoubleSpherRes}, which is symmetric in angle $\varphi$ but has broken symmetry in the variable $\vartheta$, TM oscillations are necessarily homogeneous over azimuth angle. Equation \eqref{u_m-eq2} for such oscillations can be reduced to the form similar in structure to Eq.~\eqref{v_m-eq2}, namely,
\begin{align}
\label{u_0-diff_eq}
  \left[\hat{\Delta}_0 + k^2\overline{\varepsilon}-\hat{V}^{(r)}-\hat{V}^{(\varepsilon)}-\hat{V}^{(\vartheta)}\right]
  \frac{\partial u_0(r,\vartheta)}{\partial\vartheta}=0\ ,
\end{align}
where
\begin{equation}\label{V^(theta)-2}
  \hat{V}^{(\vartheta)}=\frac{\partial\ln\varepsilon(r,\vartheta)}{\partial\vartheta}\cdot
  \frac{1}{r^2}\left(\cot\vartheta +\frac{\partial}{\partial\vartheta}\right)\ .
\end{equation}

Thus, the problem of the spectrum of EM oscillations in the double-spherical resonator shown in Fig.~\ref{DoubleSpherRes} is reduced to finding the spectrum of two-dimensional operators in square brackets of Eqs.~\eqref{v_m-eq2} and \eqref{u_0-diff_eq}. These equations are identical in functional structure, being different only in the set of effective potentials entering into them. With this in mind, we will consider below the equations not for the Debye potentials, but rather for the corresponding Green functions. After going to the Fourier representation with respect to the two remaining variables, $\vartheta$ and $r$, each of these equations can be written in the universal form as a set of coupled equations for the elements of Green matrix composed of its mode Fourier components. It is convenient to represent these equations in the form
\begin{equation}\label{GF_eqs-schemat}
  \big(k^2\overline{\varepsilon}-k_{\bm{\mu}_m}^2
  -\mathcal{V}_{\bm{\mu}_m}\big)G_{\bm{\mu}_m\bm{\mu}'_m}
  -\sum_{\bm{\nu}_m\neq\bm{\mu}_m}
  \mathcal{U}_{\bm{\mu}_m\bm{\nu}_m}G_{\bm{\nu}_m\bm{\mu}'_m}=
  \delta_{\bm{\mu}_m\bm{\mu}'_m}\ ,
\end{equation}
where sub-index $m$ of the previously introduced vector mode indices corresponds to the fixed (specifically, $m$-th) azimuthal component,
\begin{equation}\label{Pert_oper_matr_els}
  \mathcal{U}_{\bm{\mu}\bm{\nu}}=
  \int_{\Omega_{\mathrm{out}}}\!\!\!d\mathbf{r}\,
  \bra{\mathbf{r};\bm{\mu}}\hat{V}\ket{\mathbf{r};\bm{\nu}}
\end{equation}
is the element of zero-diagonal matrix of \emph{inter-mode} potentials, since $\bm{\nu}\neq\bm{\mu}$, and $\mathcal{V}_{\bm{\mu}}\equiv\mathcal{U}_{\bm{\mu}\bm{\mu}}$ is the \emph{intra-mode} potential.

The infinite set of equations \eqref{GF_eqs-schemat} can be solved with the aid of the method of separation of the resonator modes in inhomogeneous systems, which is detailed in Ref.~\cite{bib:GanapErTar07}. According to this method, all the inter-mode (non-diagonal) propagators in \eqref{GF_eqs-schemat} can be linearly expressed in terms of the corresponding diagonal propagators via formula
\begin{equation}\label{Gnu->Gmumu_m}
  G_{\bm{\nu}_m\bm{\mu}_m}=\bra{\bm{\nu}_m}
  (\openone-\hat{\mathsf R}_m)^{-1}\hat{\mathsf R}_m
  \ket{\bm{\mu}_m}G_{\bm{\mu}_m\bm{\mu}_m}\ .
\end{equation}
Here, $\hat{\mathsf{R}}_m=\hat{\mathcal{G}}^{(V)}_m\hat{\mathcal{U}}_m$ is the mode-mixing operator defined on the whole space of mode indices, $\hat{\mathcal{G}}^{(V)}_m$ is the trial operator propagator the elements of which are the trial mode Green functions
\begin{equation}\label{Trial_G_mu_m}
  G_{\bm{\mu}_m}^{(V)}  = \left( k^2\overline{\varepsilon}-k_{\bm{\mu}_m}^2 - {\cal V}_{\bm{\mu}_m} \right)^{ - 1}
\end{equation}
which do not account for inter-mode scattering; $\hat{\mathcal{U}}_m$ is the operator inter-mode potential whose matrix is composed of elements $\mathcal{U}_{\bm{\nu}_m\bm{\nu}'_m}$ with $\bm{\nu}_m\neq\bm{\nu}'_m$, being thus zero-diagonal. For diagonal propagators $G_{\bm{\mu}_m\bm{\mu}_m}$ entering Eq.~\eqref{Gnu->Gmumu_m} we obtain unrelated equations, from which the result is obtained
\begin{equation}\label{Gmumu}
  G_{\bm{\mu}_m\bm{\mu}_m}=\left(k^2\overline{\varepsilon}-k^2_{\bm{\mu}_m}-{\mathcal
  V}_{\bm{\mu}_m}-{\mathcal T}_{\bm{\mu}_m}\right)^{-1}\ .
\end{equation}
Here
\begin{equation}\label{T-matr}
  {\mathcal T}_{\bm{\mu}_m}=\bra{\bm{\mu}_m}\hat{\mathcal{U}}_m
  (\openone-\hat{\mathsf{R}}_m)^{-1}\hat{\mathsf{R}}_m\ket{\bm{\mu}_m}
\end{equation}
is the part of the eigen-energy of the mode $\bm{\mu}_m$ which is associated exclusively with the scattering between \emph{different} resonator modes (so called mode \emph{T}-matrix).

The expression \eqref{Gmumu} for the mode propagator, which we have obtained by way of strict separation of the resonator modes in the equation set \eqref{GF_eqs-schemat}, completely determines the Green's function of the wave operator, and thus all the dynamic properties of the resonator under investigation. Using this formula, one can easily find the resonator spectrum, since for quantum as well as wave systems the spectrum is by definition determined by the maxima of the density of states (DOS), which is related to the Green operator by formula \cite{bib:Economou06}
\begin{align}\label{Dens_states->G}
  \nu(k)=\pi^{-1}\mathrm{Im}\big\{\Tr\hat{\mathcal{G}}^{(+)}\big\}\ .
\end{align}
Here, index (+) means the retarded nature of the operator, i.\,e., the sign of the imaginary part of its eigen-energy. In the next section, using the above-obtained formulas, we proceed with numerical analysis of the spectrum of the resonator depicted in Fig.~\ref{DoubleSpherRes}.

\section{Numerical experiment}
\label{Numerix}

\begin{figure}[!h]
  \setcaptionmargin{.2in}%
  \centering
  \scalebox{.6}[.6]{\includegraphics{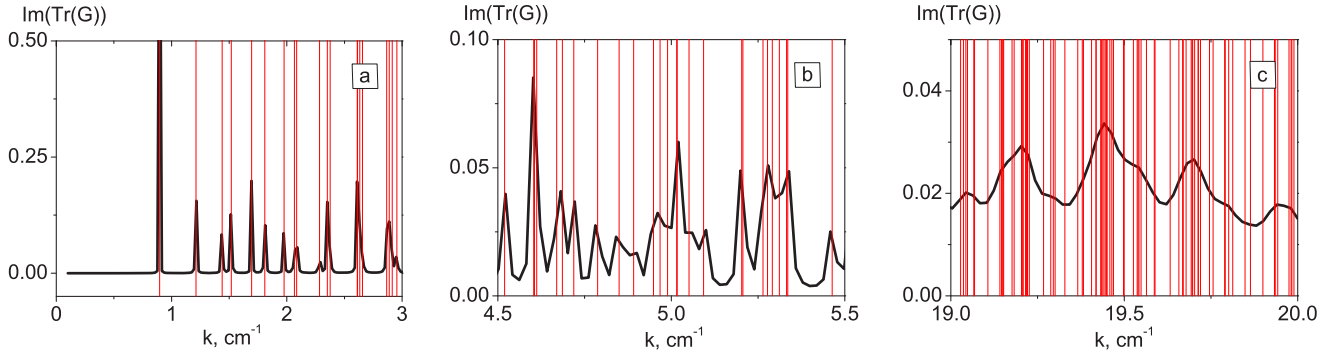}}
  \captionstyle{flushleft}\caption{The $\pi\times$density of states (black curves) and matrix $\{\hat{\mathcal{G}}^{-1}\}_{\bm{\mu}\bm{\nu}}$ eigenvalues (red lines) for the centered-layered spherical cavity resonator in three different parts of the frequency axis. The number of basis modes equals 1000, the resonator parameters are as follows: $R_{\mathrm{in}}=2.0$~cm, $R_{\mathrm{out}}=4.46$~cm, $\varepsilon_{\mathrm{in}}=2.08+10^{-4}i$, $\varepsilon_{\mathrm{out}}=1.0+10^{-4}i$.}
 \label{Spectrum}
\end{figure}

To find the spectrum directly from Green function expression \eqref{Gmumu} with exclusively analytical methods seems to be an awkward procedure. Thus, for this to be carried out to the full extent some numerical methods should be also utilized. In this regard, some essential points should be noted to be taken into account. To find the spectrum numerically from formula~\eqref{Gmumu} requires significant computational resources due to the complex structure of the intermode mass operator $\mathcal{T}_{\bm{\mu}_m}$. Therefore, if there is no need to determine such spectral characteristics as the resonance line widths or the overlap of the resonances, direct computation of the eigenvalues of matrix $\{\hat{\mathcal{G}}^{-1}\}_{\bm{\mu}\bm{\nu}}$ seems to be preferable. Quite naturally, both methods result in the same values of resonant frequencies, as one can see in Fig.~\ref{Spectrum}, yet matrix eigenvalues computing is more effective in terms of the processor time consumption.

The additional simplification due to the latter method is that it requires a smaller size of the expansion basis set (see Fig.~2b). The above presented theory formally allows for the intermode scattering between an \emph{infinite} number of basis modes. In numerical simulation, however, it is necessary to restrict oneself to a reasonable number of modes, which would allow to obtain the result with reasonable accuracy on available computing resources. Numerically computed spectrum of the resonator strongly depends on the size of the basis used for the expansions. With gradual increase in the number $N_b$ of the basis modes, an~extremely unpleasant phenomenon is observed, namely, the appearance of spurious unphysical resonances, which disappear with a further increase in the basis size. And also, some unregulated shifts of the resonance frequencies occur, which asymptotically disappear with further growing the number of basis modes. Thus, we find that it is important to determine the frequency range for which the obtained numerical result can be considered as reliable under pre-chosen size of the basis set.

The above statements are illustrated in Fig.~\ref{Spectrum}. The relatively low-frequency graphs in Fig.~\ref{Spectrum}a demonstrate  good agreement between resonance frequencies obtained from the imaginary part of the Green function trace (black curve) and the frequencies found from eigenvalues of matrix $\{\hat{\mathcal{G}}^{-1}\}_{\bm{\mu}\bm{\nu}}$ (red lines) for the case of spherical resonator with centralized dielectric insert. In Fig.~\ref{Spectrum}b, the interval of $k$ values is presented where the coincidence between resonance positions obtained in two above specified ways for the particular number of basis modes, which is chosen to be equal to $10^3$, is violated. From Fig.~\ref{Spectrum}c one can also see that with further increase in the value of wave number $k$ the coincidence totally disappears. Thus, we arrive at the conclusion that for a given number of basis modes there exists a limit value $k_{\mathrm{max}}$ below which the resonance frequencies obtained from the humps in the DOS (black curves in Fig.~\ref{Spectrum}) and the frequencies found from the inverse Green function matrix can be considered to coincide. The frequency interval from 0 to $\omega_{\mathrm{max}}$ may be thought of as the confidence range for the construction of frequency spectrum statistical characteristics for the resonator under investigation.

There is also a certain difficulty in the evaluation of intermode-scattering integrals when calculating the scattering probabilities. For high-frequency modes, the integrands are rapidly oscillating functions, and the integral values appear to be very small. This point, as well as the considerable difficulty in calculation of Bessel functions for large (especially complex) arguments, rounding errors and so on, lead to the fact that for the same relative accuracy the overwhelming part of the processor time is spent for the calculation of the off-diagonal matrix elements, a large number of which are negligibly small in absolute value and practically do not affect the spectrum.

Our investigation shows that a significant increase in the productivity with preservation of the accuracy can be achieved using the CQUAD integration algorithm \cite{bib:Connet10}, with separate control of the accuracy when integrating the diagonal and off-diagonal matrix elements. Moreover, for the off-diagonal elements it is favorable to control the absolute accuracy of the numerical integration rather than the relative accuracy. This algorithm requires more computations than the widely used algorithms of the QUADPACK package \cite{bib:Piessens83}, yet it provides the better accuracy and convergence in complex cases. In this algorithm, for each integration interval, the Clenshaw-Curtis quadratures~\cite{bib:Clenshaw60} of increasing order are calculated. The calculation error is controlled by computing the difference in sequential quadrature calculations with the integration interval bisection running on until the required computation accuracy is achieved~\cite{bib:Connet10}.

\section{Concluding remarks}
\label{Concl}

We have studied the spectrum of electromagnetic oscillations in the inhomogeneous and asymmetric spherical layered cavity resonator. In order to obtain the resonant frequency spectrum, we have developed a theoretical method for solving Maxwell's equations in inhomogeneous and asymmetric cavity resonators, which utilizes two scalar potentials only. Our study has shown that the resonance frequencies obtained from exact Maxwell equations solution for the central-symmetric layered spherical resonator coincide with matrix eigenvalues and with Green function resonances only for some initial part of the frequency axis. The length of this part depends substantially on the number of seed resonant modes that are used in our method. To increase the coincidence part, it is necessary to take the larger number of these resonant modes.

The developed theoretical method allows us, in the subsequent study, to find the frequency spectra of any asymmetric or inhomogeneous layered spherical resonator in order to determine chaos signatures in such resonant structures.

\section*{ACKNOWLEDGEMENTS}

Computer simulations were performed with the use of the grid-cluster of the Institute for Radiophysics and Electronics of the National Academy of Sciences of Ukraine, being a part of the Ukrainian National Grid infrastructure~(UNG).


\end{document}